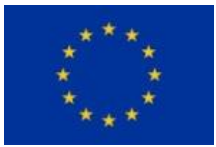
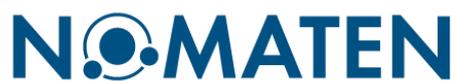

This work was carried out in whole or in part within the framework of the NOMATEN Centre of Excellence, supported from the European Union Horizon 2020 research and innovation program (Grant Agreement No. 857470) and from the European Regional Development Fund via the Foundation for Polish Science International Research Agenda PLUS program (Grant No. MAB PLUS/2018/8), and the Ministry of Science and Higher Education's initiative "Support for the Activities of Centers of Excellence Established in Poland under the Horizon 2020 Program" (agreement no. MEiN/2023/DIR/3795).

# Facet Specific Electron Conduction in Pentavalent (W$^{5+}$) WO$_3$ Drives Superior Photocatalytic CO$_2$ Reduction in (002) Plane

Muhammad Rizwan Kamal[a], Mohammad Z. Rahman[b], Amil Aligayev[c,d], Min Liu[e], Li Zhong[e], Pengfei Xia[f], Yueheng Li[f], Yue Ruan[a], Xia Xiang[a], Pir Muhammad Ismail[a], Qaisar Alam[a], Ahmed Ismail[a], Muhammad Zahid[a], Xiaoqiang Wu[g], Abdullah N. Alodhayb[h], Qing Huang[c], Raj Wali Khan[i], Fazal Raziq[b*], Sharafat Ali[b*], Liang Qiao[a*]

[a]School of Physics, University of Electronic Science and Technology of China, Chengdu 610054, China

[b]Institute for Advanced Study, Chengdu University, Chengdu 610106, China

[c]Science Island Branch of Graduate School, University of Science and Technology of China, Hefei, 230026, China

[d]NOMATEN Centre of Excellence, National Centre for Nuclear Research, 05-400 Swierk/Otwock, Poland

[e]SEU-FEI Nano-Pico Center, Key Laboratory of MEMS of Ministry of Education, Southeast University, Nanjing 210096, China

[f]Yangtze Delta Region Institute (Huzhou), University of Electronic Science and Technology of China, Huzhou 313001, P.R. China

[g]School of Mechanical Engineering, Chengdu University, Chengdu 610106, China

[h]Department of Physics and Astronomy, College of Science, King Saud University, Riyadh 11451, Saudi Arabia

[i]National Water and Energy Center, United Arab Emirates University, Al Ain 15551, United Arab Emirates

Corresponding authors e-mail: fazal.raziq@uestc.edu.cn; chemistry2827@gmail.com; liang.qiao@uestc.edu.cn

## Abstract

This article reports a concept of heat-induced topological modifications of non-layered WO$_3$ followed by successful synthesis of oxygen-vacant more-porous nanosheets with exposed active (002) facet. Experimental measurements and Density Functional Theory (DFT) calculations have revealed that the photoexcited electrons are found to accumulate preferentially on (002) facet to yield enhanced electron conduction, and consequently, strengthen the reduction potential as active catalytic sites for photocatalytic $CO_2$ reduction. Owing to these beneficial properties, the more-porous nanosheets of $WO_3$ with (002) facet have exhibited superior performance than that of less-porous nanosheets of $WO_3$ with (220) facet and bulk $WO_3$ with (205) facet. This study therefore provides a new understanding of regulating physical, optical, and electronic properties through intricate atomic structure modulation of $WO_3$, and may find widespread application in optoelectronics, sensors, and energy conversion.

## 1. Introduction

The $CO_2$ is an unambiguous source for adverse climate change and anthropogenic threat [1, 2]. Catalytic reduction of $CO_2$ into environmentally benign products offers a 'foe-to-friend' solution to address these global challenges [3-5]. Owing to exploiting solar energy directly, photocatalytic reduction of $CO_2$ has been emerged as a promising means of converting $CO_2$ into fuels and commodity chemicals [4]. However, the viability of this technique critically would depend on the earth abundant, cheap, efficient, and durable photocatalysts [6-8]. Oxides of transition metals have shown promises in this regard [4, 9]. Although great progress had been witnessed over the past few years, the past research efforts were predominantly focused on wide band gap semiconductors (e.g. $TiO_2$ [10], $SrTiO_3$ [11], $ZnO$ [12], etc.) that work only under UV light excitation. Therefore, developing active photocatalysts in visible light is a prime priority [13-15].

Narrow bandgap (2.5 - 2.8 eV) tungsten trioxide ($WO_3$) is a promising candidate for being a visible light active photocatalyst [16, 17]. However, the relatively less negative conduction band (CB) edge of pristine $WO_3$ limits its ability to efficiently drive $CO_2$ reduction reactions to products like $CH_4$ or CO. This limitation arises because the CB potential is not sufficiently negative to provide the required thermodynamic driving force for these specific reduction pathways [18-20]. Different strategies such as surface modification, heterojunction, doping and tuning morphology, etc. were adopted but the catalytic performance was marginally increased [21-23]. According to reported literature, mesoporous $TiO_2$ and titanium-based perovskite oxides, including $BaTiO_3$ and $SrTiO_3$, have been effectively designed with copper particles and oxygen vacancies to improve photocatalytic $CO_2$ reduction. For example, $TiO_2$ decorated with different quantities of Cu particles has a considerable effect on the photocatalytic reduction of $CO_2$ into $CH_4$ and CO [24]. Furthermore, improved catalytic activity is demonstrated by an $Ag/TiO_2$ hybrid that is enriched in oxygen vacancies, leading to increased yields of $CH_4$ and CO [25]. When compared to pure $TiO_2$, the introduction of Cu in an ideal composite ($TiO_2$-Cu, 5% mole ratio) increases the yield of $CH_4$ by 2.2 times and the yield of CO by 3 times [26]. Additionally, during light irradiation, $TiO_2/Cu_2O$ composite nanorods exhibit enhanced photocatalytic $CO_2$ reduction. Interestingly, after 4 hours of irradiation, the $TiO_2/Cu_2O$-15% sample shows the highest $CH_4$ output of 1.35 $\mu mol\ g^{-1}\ h^{-1}$ , which is 15 times higher than that of $Cu_2O$ nanoparticles and 3.07 times higher than that of pristine $TiO_2$ nanorods [27]. Further, it has been reported that the ZnO and $TiO_2$ with particular exposed facets can effectively enhance the catalytic performance[28-31] due to altered surface atomic configuration and local chemical environment that significantly influence the heterogeneous reactivity [16, 32]. The crystal facets were shown to play crucial roles in photocatalysts for $CO_2$ activation and subsequent reduction reactions [33-35]. Inspired by above studies, we attempted to synthesize nanosheets of $WO_3$ with defined facets. Recent studies have highlighted that oxygen vacancies (OVs) play a pivotal role in the photocatalytic reduction of $CO_2$ by serving as active electron-rich sites that facilitate $CO_2$ adsorption and subsequent activation. OVs are known to modify the local electronic structure, creating defect states near the conduction band edge, which enhance the interaction with $CO_2$ intermediates such as COOH, CO, and $CH_4$*. These defect sites also lower the activation barriers for key reduction pathways, enabling selective and efficient photocatalytic conversion of $CO_2$ into valuable products.

However, the synthesis of nanosheets from non-layered $WO_3$ is a daunting task because of difficulties in bond cleaving and the lack of intrinsic driving forces for 2D anisotropic growth [36, 37]. Furthermore, challenges are lied in the intrinsic thermodynamic instability of facets in monoclinic $WO_3$ structures. We here report a strategy of fast heat-induced topological transformation to synthesize oxygen-vacancy (OVs) rich more-porous $WO_3$ nanosheets with stabilized (002) crystal facets. Experimental results show that these OVs and pore-rich $WO_3$ nanosheets with exposed facets positively impact the charge transport and catalytic kinetics.

The plane of exposed crystal facet and OVs were evaluated using X-ray diffraction (XRD), transmission electron microscopy (TEM), X-ray photoelectron spectroscopy (XPS), Raman spectroscopy, and electron paramagnetic resonance spectroscopy (EPR). $CO_2$ temperature-programmed desorption (TPD) and density functional theory (DFT) calculations were utilized to investigate and monitor $CO_2$ activation, adsorption, and chemical evolution on the surface of $WO_3$ during the photocatalytic process. This study provides comprehensive insights into how oxygen vacancies on the (002) facet of $WO_3$ contribute to enhanced electron conduction, intermediate stabilization, and selective product formation, offering a mechanistic understanding of their critical role in $CO_2$ photoreduction.

As discussed elaborately in the following sections, it has been found that the existence of OVs might modify the electrical band structure of $WO_3$ in addition to facilitating the charge transport kinetics but also significantly improve the catalytic activation of $CO_2$ molecules, thereby boosting the $CO_2$ reduction efficiency. This study therefore opens a new direction to pursue $WO_3$ as a promising candidate for $CO_2$ reduction in semiconductor solid state materials for solar fuel production.

## 2. Experimental section

All products (99.9% purity) were obtained from Beijing Chemical Co. Ltd.

### 2.1. Synthesis and Liquid Exfoliation of Bulk $WO_3 \cdot 2H_2O$

Firstly, bulk $WO_3 \cdot 2H_2O$ was prepared by dissolving 200 mg of Sodium tungstate dihydrate ($Na_2WO_4 \cdot 2H_2O$) in 150 mL of 4.8 M nitric acid ($HNO_3$) and stirring for 72 hours at 20°C. We got yellow precipitates and were separated, later washed with deionized (DI) water, and dried overnight at 80°C. To exfoliate the bulk $WO_3 \cdot 2H_2O$ into nanosheets, 10 mg of the bulk $WO_3 \cdot 2H_2O$ was dissolved in 20 mL of DI water and ultrasonicated in an ice bath for 3 hours. The pale-yellow suspension was then centrifuged to remove unsuspended particles and subsequently dried at 60°C overnight, yielding $WO_3 \cdot 2H_2O$ nanosheets.

### 2.2. Synthesis of more-porous and less-porous $WO_3$ nanosheets

More-porous $WO_3$ nanosheets were synthesized from $WO_3 \cdot 2H_2O$ nanosheets by a "topological modification" by calcining at 400°C for 0.5 hours. Less-porous $WO_3$ nanosheets were produced using the same method, by setting the temperature at 500°C for 10 minutes. The structural alterations at the atomic and nanoscale levels that take place during calcination and impact the accessibility and connectivity of the pores are referred to as "topological modification". For comparison, bulk $WO_3$ was obtained by calcining the bulk $WO_3 \cdot 2H_2O$ precursor at 500°C for 3 hours.

## 3. Physical Characterization Techniques

A Bruker D8 advanced diffractometer equipped with Cu K radiation was used to collect XRD diffraction patterns. A Shimadzu UV2700 spectrophotometer was utilized to measure the UV-visible diffuse reflectance spectra (DRS) of $BaSO_4$. A transmission electron microscope, specifically a JEOL JEM-2010 operating at 200 kV, was used for imaging. Photoluminescence (PL) spectra were acquired using a Hitachi F-4500 fluorescence spectrophotometer.

The time-resolved photoluminescence (TRPL) spectra were obtained on an FLS1000 fluorescence lifetime spectrophotometer (Edinburgh, Instruments, UK) using 370 nm-excitation lights. The decay curves of $(205)WO_3$, $(220)WO_3$, and $(002)WO_3$ can be fitted by the tri-exponential equation below:

$$y = A_1 \times exp(-x/t_1) + A_2 \times exp(-x/t_2) + A_3 \times exp(-x/t_3) + y_0$$

in which $y_0$ indicates y offset, $A_1$, $A_2$ and $A_3$ represent the amplitude and $t_1$, $t_2$ and $t_3$ represents decay constants. The average lifetime ($\tau_{ave}$) is calculated by the following equation:

$$\tau_{ave} = (A_1 \times t_1^2 + A_2 \times t_2^2 + A_3 \times t_3^2) \div (A_1 \times t_1 + A_2 \times t_2 + A_3 \times t_3)$$

## 4. Results and discussion

We synthesized bulk, less-porous, and more-porous $WO_3$ according to the protocol shown in **Scheme 1**, In a typical experiment, $Na_2WO_4 \cdot 2H_2O$ and $HNO_3$ were used to obtain the layered $WO_3 \cdot 2H_2O$. Bulk-$WO_3$ was obtained by directly calcining layered $WO_3.2H_2O$ at 500 °C for 30 min. The microstructure and morphology of the as-prepared samples were assessed via scanning electron microscopy (SEM) and transmission electron microscopy (TEM). The SEM images show sheet-like morphology as shown in **Fig. S1a-c** and TEM images of the directly calcined sample illustrate a bulk morphology, whereas the exfoliated products exhibit a distinct nanosheet-like morphology **(Fig. S2-S4)**. Particularly, the calcined product derived from $WO_3.2H_2O$ nanosheets at 400°C for 30 minutes distinctly demonstrates a large-area 2D nanosheet-like more-porous structure (**Fig. 1a**). In the respective high-resolution TEM (HRTEM) images (**Fig. 1b-d**), the observed lattice fringes can be confidently attributed to the (205), (220), and (002) planes, providing strong evidence of the exposed lattice structure of $WO_3$. Importantly, the marked red circles in **Fig. 1d** indicate the incurred surface defects within the (002)$WO_3$ structure due to the higher density of OVs than that of (205)$WO_3$ and (220)$WO_3$. It was further substantiated by the analysis of the inverse fast Fourier transform (FFT) patterns. Specifically, the region highlighted by red circles in the inset of **Fig. 1d** reveals an increased level of distortion in the inverse FFT pattern of (002)$WO_3$, indicating the presence of a greater number of defects compared to (205)$WO_3$ and (220)$WO_3$ (insets of **Fig. 1b, c**), respectively.

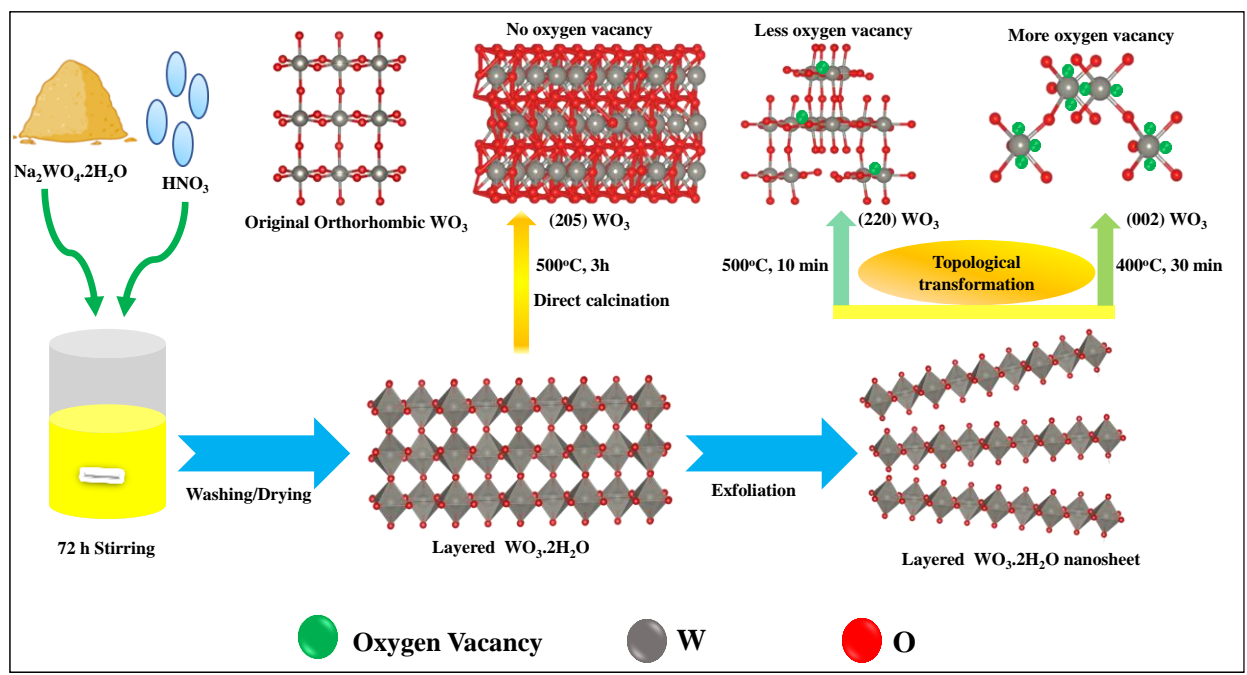

**Scheme 1.** Schematic illustration of the rational design and synthesis process of different facets of $WO_3$.

The selected area electron diffraction patterns (SAED) of $WO_3$ crystal planes, specifically (205), (220), and (002), as depicted in **Fig. S5**, reveal notable distinctions and confirm the presence of relevant facets that further support our claim. Particularly, the (002)$WO_3$ pattern exhibits a greater number of diffraction spots, indicative of an augmented d-spacing value. Moreover, compared to the (205)$WO_3$ and (220)$WO_3$ samples, the (002)$WO_3$ sample showcases more irregular fringes, suggesting a high prevalence of oxygen vacancies and structural distortion within its crystal lattice [39] which is also evident in the obtained X-ray diffraction (XRD) patterns as shown in **Fig. S6**. The diffraction peaks observed in the XRD patterns for (205)$WO_3$, (220)$WO_3$, and (002)$WO_3$, respectively, are in accordance with the standard pattern of orthorhombic WO₃ (JCPDS Card No. 96-210-7313), verifying the high purity of the samples. The purity of the samples is also evident by identical Raman peaks position at almost the same position (**Fig. 1e**). However, the diffraction planes of (002)WO₃ were broader and weaker than those of (220)WO₃ and (205)WO₃, indicating a higher degree of lattice distortion in (002)WO₃, which corresponds to the increased existence of defects within the crystal structure [40]. Defects in crystals disrupt the regular arrangement of atoms, leading to imperfections that result in a decrease in crystal size [41]. The crystallite sizes of all three catalysts were measured using the Scherrer formula.

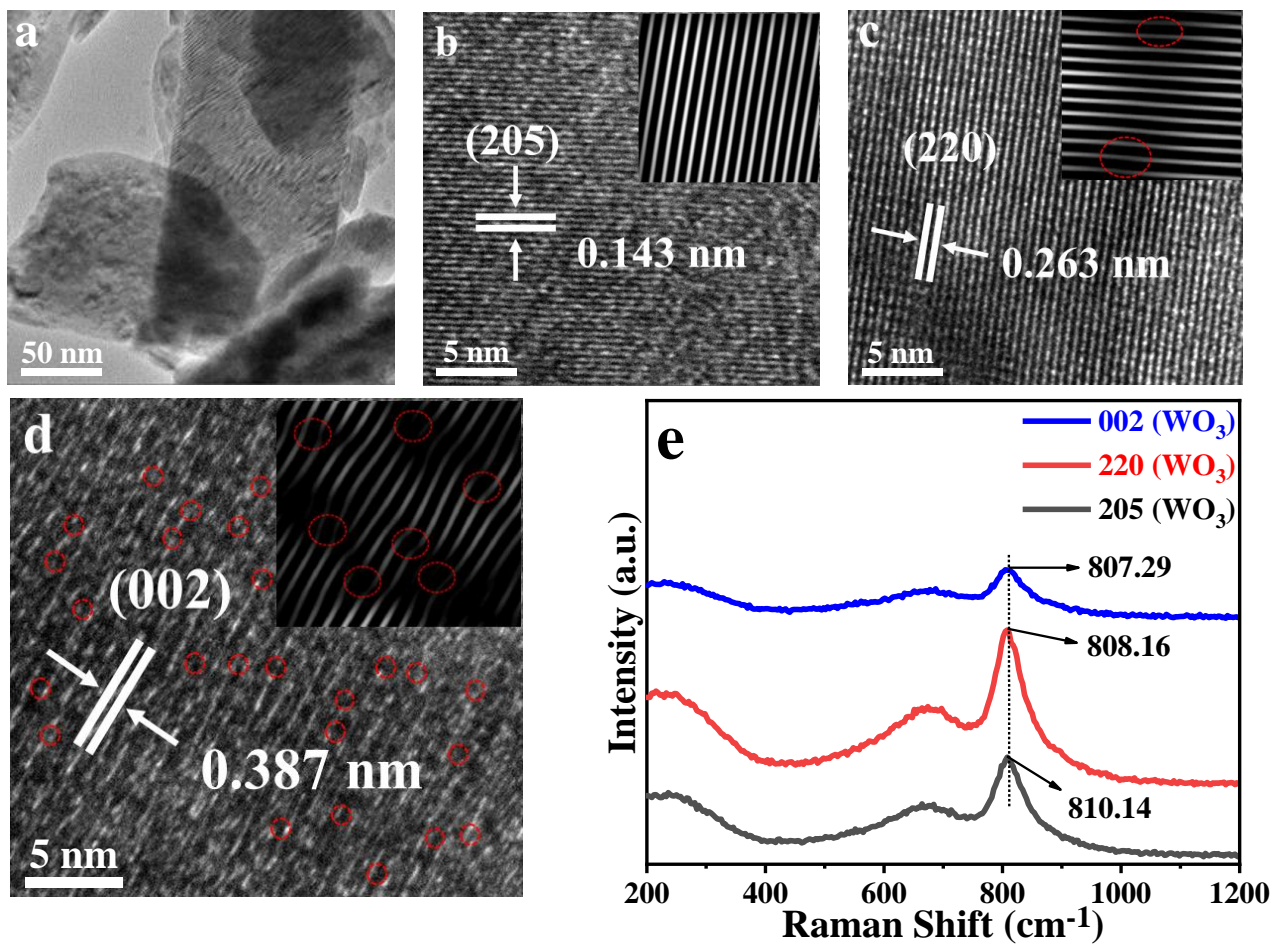

**Figure 1. Confirmation of the formation of defined facets and altered physical structure in WO₃.** (a) TEM image of (002)WO₃, (b-d) HR-TEM, IFFT Pattern (inset) of (205)WO₃, (220)WO₃, and (002)WO₃, and (e) Raman spectroscopy.

The crystal size of (002)WO₃ (16. 77 nm) was less than that of (220)WO₃ (39.54 nm) and (205)WO₃ ( 40.61 nm), as listed in **Table 1**. Correspondingly, the surface area was evaluated using $N_2$ adsorption-desorption isotherms, revealing type-IV isotherms with distinct H3-type hysteresis loops for each catalyst. The measured surface areas for $(205)WO_3$, $(220)WO_3$, and $(002)WO_3$ were 17.48, 21.74, and 24.10 $m^2 \cdot g^{-1}$, respectively, **(Fig. S7-9)** while their pore volumes are 0.08, 0.10, and 0.12 $cm^3 \cdot g^{-1}$ **(Table 1)**, demonstrating that $(002)WO_3$ exhibits higher surface area that is beneficial for enhancing photocatalytic efficiency [42]. The X-ray photoelectron spectroscopy (XPS) was exploited to investigate the surface chemical state and chemical composition of the as-prepared samples. The XPS survey spectrum **(Fig. 2a)**, confirms the coexistence of the tungsten (W) and oxygen (O) peaks in $(205)WO_3$, $(220)WO_3$, and $(002)WO_3$, which is in agreement with the Energy Dispersive X-ray Spectroscopy (EDX) as is shown in **Fig. S2-4**.

**Table 1.** Physicochemical parameters such as surface area, pore size, pore volume, crystal size, pentavalency ratio, oxygen ratio, and charge transfer resistance ($R_{CT}$) of $(205)WO_3$, $(220)WO_3$, and $(002)WO_3$.

| Samples | BET surface Area $(m^2/g)$ [a] | Pore Size (nm) [a] | Pore Volume $(cm^3/g)$ [a] | Crystal size (nm) [b] | $W^{5+}/W$ (%) [c] | $O_V/O$ (%) [d] | $R_{CT}$ $[\Omega]$ [e] |
|---|---|---|---|---|---|---|---|
| $(205)$ $WO_3$ | 17.48 | 2.99 | 0.08 | 40.61 | 9.0 | 7.7 | 28838 |
| $(220)$ $WO_3$ | 21.74 | 3.11 | 0.11 | 39.54 | 13.4 | 10.2 | 14518 |
| $(002)$ $WO_3$ | 24.11 | 3.12 | 0.12 | 16.77 | 23.0 | 17.2 | 6665 |

[a]. Measured from $N_2$ adsorption-desorption isotherms. [b]. Crystallite size calculated by XRD using Scherrer equation. [c]. Ratio of $W^{5+}$ to $(W^{5+} + W^{6+})$. [d]. Ratio of $O_V$ to $(O_{latt} + O_{ads} + O_V)$. [e]. charge transfer resistance $(R_{CT})$.

The high-resolution XPS spectra for W 4f and O 1s are respectively depicted in **Fig. 2b-c**. The photoelectrons produced by the spin-orbit splitting of the $4f_{7/2}$ and $4f_{5/2}$ of $W^{6+}$ species, respectively, are responsible for the binding energies of W 4f at 35.64 and 37. 79 eV in $(205)WO_3$, 35.65 and 37.8 eV in $(220)WO_3$, 35.72 and 37.84 eV in $(002)WO_3$, and the weak peaks at 35.2 and 37.3 eV in $(205)WO_3$, 35.3 and 37.4 eV in $(220)WO_3$, 35.1 and 37.2 eV in $(002)WO_3$ are associated with $W^{5+}$ species [43-45]. The identification of $W^{5+}$ species is indicative of OVs formation [43, 46, 47]. Additionally, the relative proportions of $W^{5+}$ to total W across different crystal facets of $WO_3$ are quantitatively presented in **Table 1**. Notably, the $(002)WO_3$ sample exhibits a higher percentage of $W^{5+}$ compared to the $(220)WO_3$ and $(205)WO_3$ samples, suggesting variations in the chemical environment and defect structure across these facets. As depicted in **Figure 2b**, the O 1s spectra for the $(205)WO_3$, $(220)WO_3$, and $(002)WO_3$ samples can be deconvoluted into three distinct peaks. Specifically, the peaks located at 530.44, 530.34, and 531.09 eV are attributed to lattice oxygen ($O_L$) atoms within the crystal lattice, the peaks at 531.28, 531, and 531.28 are representing adsorbed oxygen ($O_{ads}$), whereas the peaks at 531.02, 530.65, and 531.12 eV are associated with oxygen vacancy ($O_v$) [48]. We have also performed the quantitative analysis of oxygen vacancy ($O_V$) to total O ratio ($O_V/O$) across different facets of $WO_3$ and the results are shown in **Table 1**. We have found that the atomic percentage (at. %) of $O_V$ in the $(002)WO_3$ sample is 17.2%, which is significantly higher than that in the $(220)WO_3$ (10.2%) and $(205)WO_3$ (7.7%) samples [38, 43]. These results unequivocally indicate that the $(002)WO_3$ sample exhibits a substantially higher concentration of surface OVs, underscoring its unique chemical surface properties.

The impacts of oxygen vacancies (OVs) was evaluated using electron paramagnetic resonance (EPR) spectroscopy, a sophisticated analytical method employed for identifying unpaired electrons in catalysts. EPR spectroscopy **(Fig. 2d)** confirms the presence of unpaired electrons associated with OVs, particularly in $(002)WO_3$, which exhibits a prominent g = 2.003 signal **(Fig. S10)**, evidently surpassing signals recorded for $(220)WO_3$ and $(205)WO_3$. These unpaired electrons enhance electron mobility and act as active centers for transferring electrons to adsorbed $CO_2$ molecules [41, 46]. Affirmatively, these findings therefore confirm the augmented electron conduction on the pentavalent $(002)WO_3$ facet.

The DFT calculations were conducted to ascertain the formation of OVs and identify facets with higher concentrations of OVs. To achieve this, the OVs formation energy was computed that was -2.13, -2.92, and -3.82 eV, respectively, for $(205)WO_3$, $(220)WO_3$, and $(002)WO_3$ **(Fig. 2e-g)**. This means that the formation of oxygen vacancies is more favourable on $(002)WO_3$ facets [9].

The light absorption and energy band structure of the samples were assessed through examination of their respective UV-visible spectra, as depicted in **Figure 3a**. Each sample exhibited the capability to absorb visible light. The $(002)WO_3$ displayed the highest intensity in visible light absorption that often observed due to the inclusion of oxygen vacancies (OVs) [49, 50].

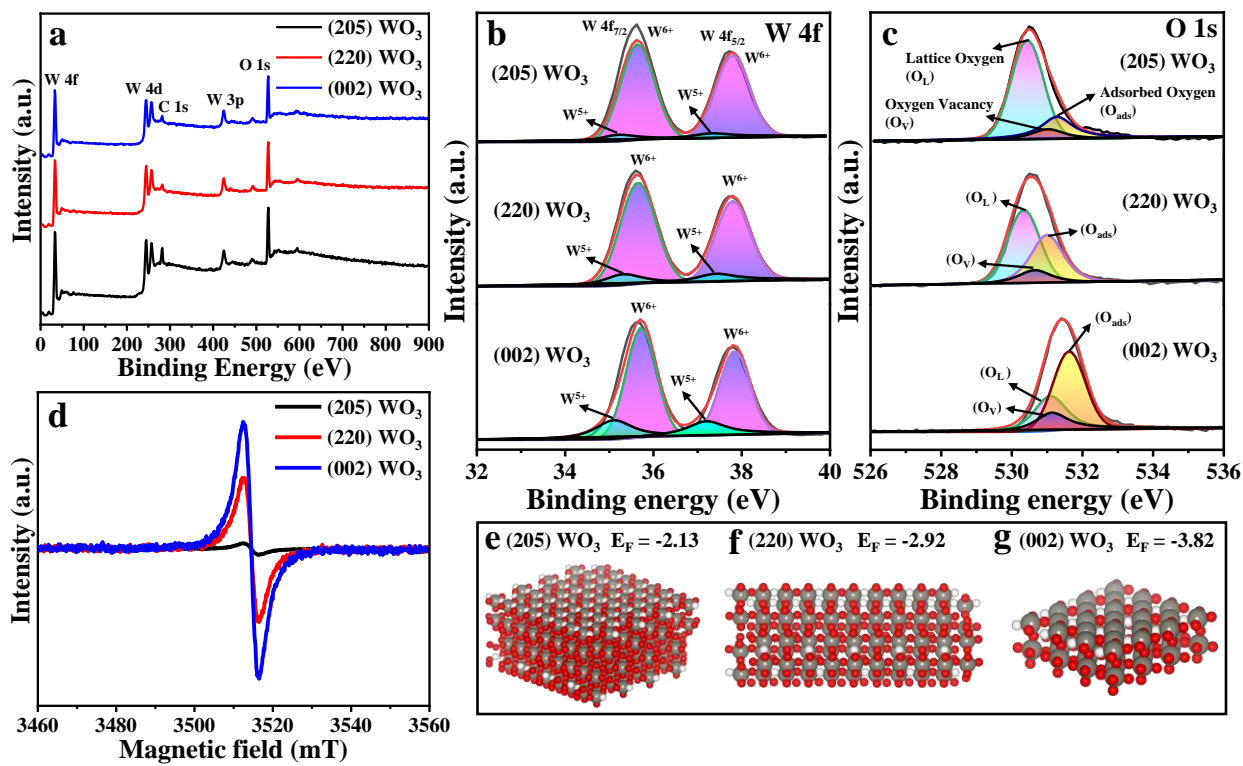

**Figure 2. Evidence of oxygen induced pentavalency and augmented electron conduction in (002) facet of WO₃.** (a) XPS survey spectra, (b-c) Deconvoluted high-resolution XPS spectra of W 4f and O 1s core level, (d) EPR Spectra, and (e-g) the vacancy formation energies of (205)WO₃, (220)WO₃ and (002)WO₃.

The corresponding band gap energies (E$_g$) of (205)WO₃, (220)WO₃, and (002)WO₃ were determined as 2.74 eV, 2.65 eV, and 2.39 eV, respectively **(Fig. 3b)**, in accordance with the Tauc formula [51, 52]. Consequently, the band structures of (002)WO₃ and (220)WO₃ can be inferred. Additionally, the observed colour variations in our WO₃ samples (yellow, dark green, and light green) are attributed to reductions in band gap **(Fig. 3c)**. Mott-Schottky (M-S) plots were carried out to investigate the band structures of (205)WO₃, (220)WO₃ and (002)WO₃ as displayed in **Fig. 3d**. The E$_{fb}$ of (205)WO₃, (220)WO₃ and (002)WO₃ can be found to be -0.04, -0.23 and -0.32 V vs. NHE, respectively, by extending the linear part of M-S plots [52-54]. The positive slope of M-S plots is seen in all investigated samples, demonstrating the characteristics of n-type semiconductors [52, 54]. According to prior reports, in n-type semiconductors, the CB potential is generally deeper than the flat-band potential (0.1−0.2 eV) [23], thus the E$_{CB}$ of (205)WO₃, (220)WO₃ and (002)WO₃ can be discerned to -0.14, -0.33 and -0.42 V vs. NHE (the difference value was set to be 0.1 eV here). The steady-state PL emission spectra of three $WO_3$ samples are presented in **Fig. 3e**. The (002)$WO_3$ facet exhibits the lowest PL intensity among the (205)$WO_3$ and (220)$WO_3$ facets, indicating the reduced recombination rate of photogenerated charge carriers [52, 54, 55]. This is further confirmed by the time-resolved PL spectra (TRPL), as presented in **Fig. 3f**. TRPL decay curves and fitting with tri-exponential model for (205)$WO_3$, (220)$WO_3$, and (002)$WO_3$ is shown in **Fig. S11-13**. The calculated average PL lifetimes at the excitation wavelength of 370 nm are as follows: 0.706, 0.716, and 0.806 ns attributed to (205)$WO_3$, (220)$WO_3$, and (002)$WO_3$, respectively, representing an immensely decreased rate of recombination in (002)$WO_3$ facet [56]. Together PL and TRPL prove the photo-excited charge carrier's exceptional ability to separate and transfer in the (002)$WO_3$ sample under simulated light irradiation. Further, we carried out electrochemical transient photocurrent density measurements (i-t) and electrochemical impedance spectroscopy (EIS) to gauge the extent of charge transfer behavior [52, 54, 55]. It is obvious that the trend in the charge transfer and separation efficiency follows this order: (002)$WO_3$ > (220)$WO_3$ > (205)$WO_3$. The charge-transfer resistance is shown in **Table 1**. The lowest charge transfer resistance ($R_{CT}$) for (002)$WO_3$ suggests the highest overall conductivity of (002)$WO_3$. The enhanced electron conduction in (002)$WO_3$ is corroborated by higher photocurrent density **(Fig. 3g)** and its lower charge transfer resistance **(Fig. 3h and Table 1)**. This superior charge transport behavior arises from the synergistic effects of a higher OV concentration and the unique electronic properties of the (002) facet, as confirmed by Mott-Schottky plots and band structure analysis **(Figs. 3d and 4a–c)**. We evaluated the $CO_2$ photoreduction activity of the as-synthesized samples using gas chromatography. We detected two products, such as CO and $CH_4$ evolved upon $CO_2$ photoreduction. As seen in **Fig. S14, 15** the production rate of CO and $CH_4$ over (205)$WO_3$, (220)$WO_3$ and (002)$WO_3$ increases almost linearly with the irradiation time. The rate of CO production was ~10, ~40, and ~120 $\mu$molg$^{-1}$h$^{-1}$ for (205)$WO_3$, (220)$WO_3$, and (002)$WO_3$ respectively, while the rate of $CH_4$ production was ~4, ~24 and ~60 $\mu$molg$^{-1}$h$^{-1}$, respectively **(Fig. 3i)**. Further, it is evident from **Figs. 3j** and **S16** that the main products, $^{13}CO$ (m/z = 29) and $^{13}CH_4$ (m/z = 17), originate from the conversion of $CO_2$ rather than the decomposition of (002)$WO_3$. Confirming $CO_2$ as the carbon source for the reported products is the main goal of this interpretation, which is in line with earlier research. Comparison of the photocatalytic performances of (002)$WO_3$ with other overall reported literature on $WO_3$ is also depicted in **Table. S2**. Interestingly, the exceptional $CO_2$ conversion efficiency of (002) $WO_3$ is further achieving an apparent quantum efficiency (AQE) of approximately 0.84 % at 420 nm (See supporting information for details). Notably, this AQE is more superior than $(220)WO_3$ (0.33%), and $(205)WO_3$ (0.10%) as shown in (**Fig. 3l**). Aside from the higher photocatalytic performance, we also evaluated the photocatalytic stability of the as-prepared samples. As shown in **Fig. 3k**, the $(002)WO_3$ sample exhibited the best stability, while the $(205)WO_3$ and $(220)WO_3$ samples are presented in **Fig. S17**, where their stability data are shown. The results indicate that the as-prepared samples demonstrate excellent stability.

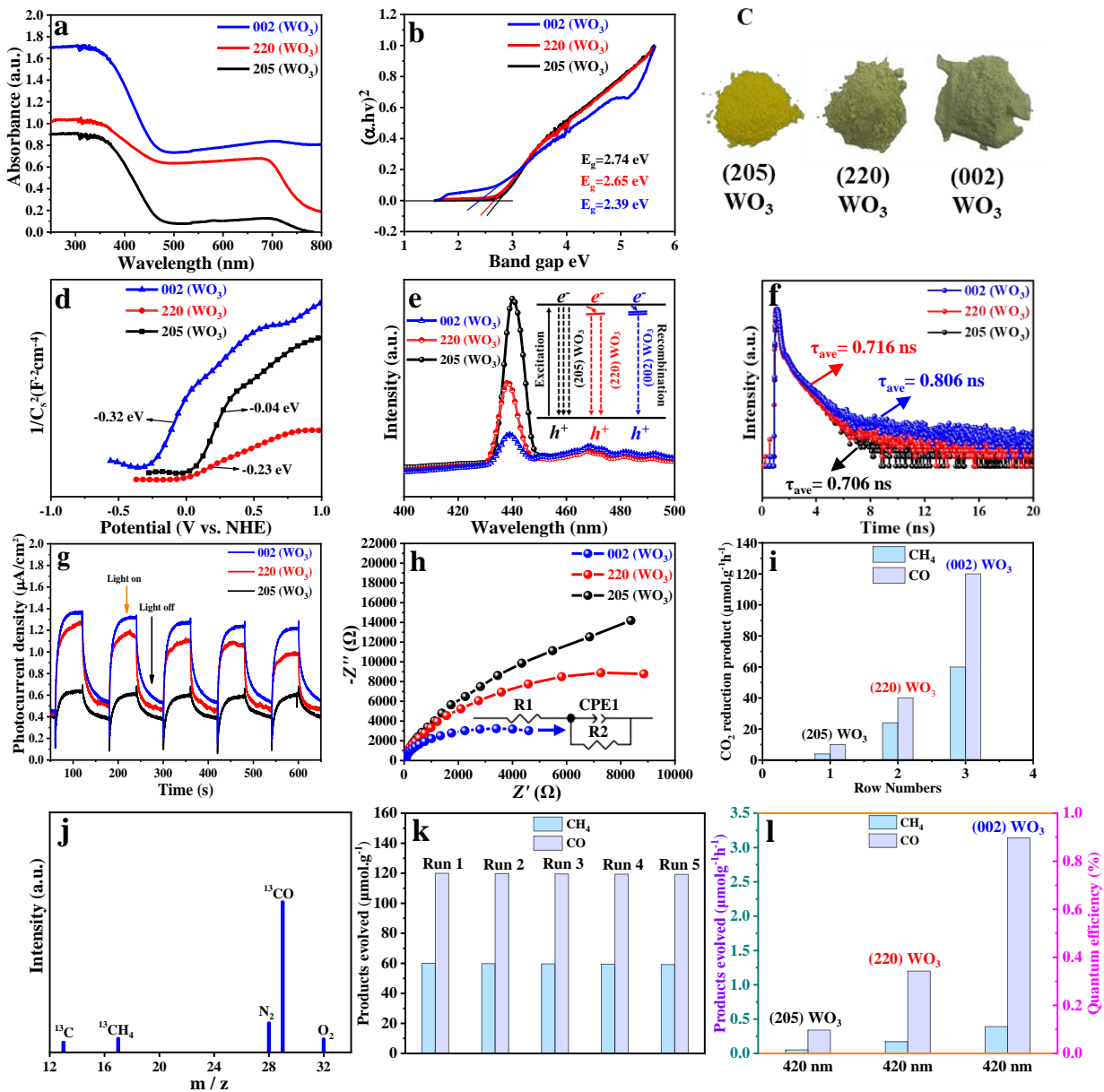

**Figure 3. Evaluation of photo-physicochemical properties and CO₂ reduction performance.** (a) UV-vis diffused reflectance spectra, (b) Tauc plots, (c) digital images of as prepared samples showing the colour variations as an indication of narrowing bandgap, (d) Mott-Schottky plots, (e) photoluminescence spectra, (f) TRPL spectra obtained at the excitation wavelength of 370, (g) electrochemical transient photocurrent, (h) Nyquist plots for electrochemical impedance spectroscopy (EIS), (i) photocatalytic CO₂ reduction, (j) Mass spectroscopy of the productions over (002)WO₃ in photocatalytic reduction of ¹³CO₂, (k) Stability test of (002)WO₃ under vis-light irradiation (420 nm ≥) for the production rate of CO and CH₄, and (l) quantum efficiency of CO₂ reduction on (205)WO₃, (220)WO₃ and (002)WO₃.

Furthermore, following prolonged use, structural investigations employing XRD (**Fig. S18**) and Raman spectroscopy (**Fig. S19**) further verify that there is almost no change after stability test. To further delve into the electronic band structure and density of states (DOS), we have carried out Density Functional Theory (DFT) calculations for $(205)WO_3$, $(220)WO_3$, and $(002)WO_3$ (**Fig. 4a-f**). We have performed a comprehensive structural optimization using standard DFT/GGA functional, and the optimized geometries are shown in **Fig. S20-22**. At room temperature, all these $WO_3$ samples feature a clearly defined band gap between their bonding and anti-bonding states and exhibit indirect bandgap energy as the valence band maximum (VBM) and conduction band minimum (CBM) occur at different G-Z k-points. The calculated bandgap energy for $(205)WO_3$, $(220)WO_3$, and $(002)WO_3$ is 2.56 eV, 2.36 eV, and 2.23 eV, respectively **Fig. 4a-c**. Although, the calculated band is slightly underestimated, but consistent with experimental results. This underestimation is a well-documented limitation of the GGA functional used in DFT calculations. The DOS analysis reveals that the valence band is predominantly populated by O 2p states and the conduction band by W 5d states. In the case of $(002)WO_3$, the valence band in the -10 to 0 eV range is predominantly shaped by O 2p electronic states, augmented by contributions from W 6s and 5d states. Conversely, the conduction band (2.23–12 eV) is primarily influenced by W 5d electron states, complemented by contributions from O 2p and W 6s states. A pronounced dispersion of bands can be observed in the conduction bands across the high-symmetry points that suggests a low effective mass for electrons. Moreover, the conduction band edge of $(002)WO_3$ is significantly shifted to the more negative potential, thereby strengthening the reduction potential of photogenerated electrons for $CO_2$ reduction. The DOS and band structure of $(002)WO_3$ reveal defect states near the Fermi level, which position the CBM close to the Fermi level and facilitate efficient electron donation to adsorbed species. These oxygen vacancies create electron-rich sites that enhance the adsorption of intermediates like CO* and COOH*, enabling their subsequent reduction to products such as CH₄* and CH₃OH*.

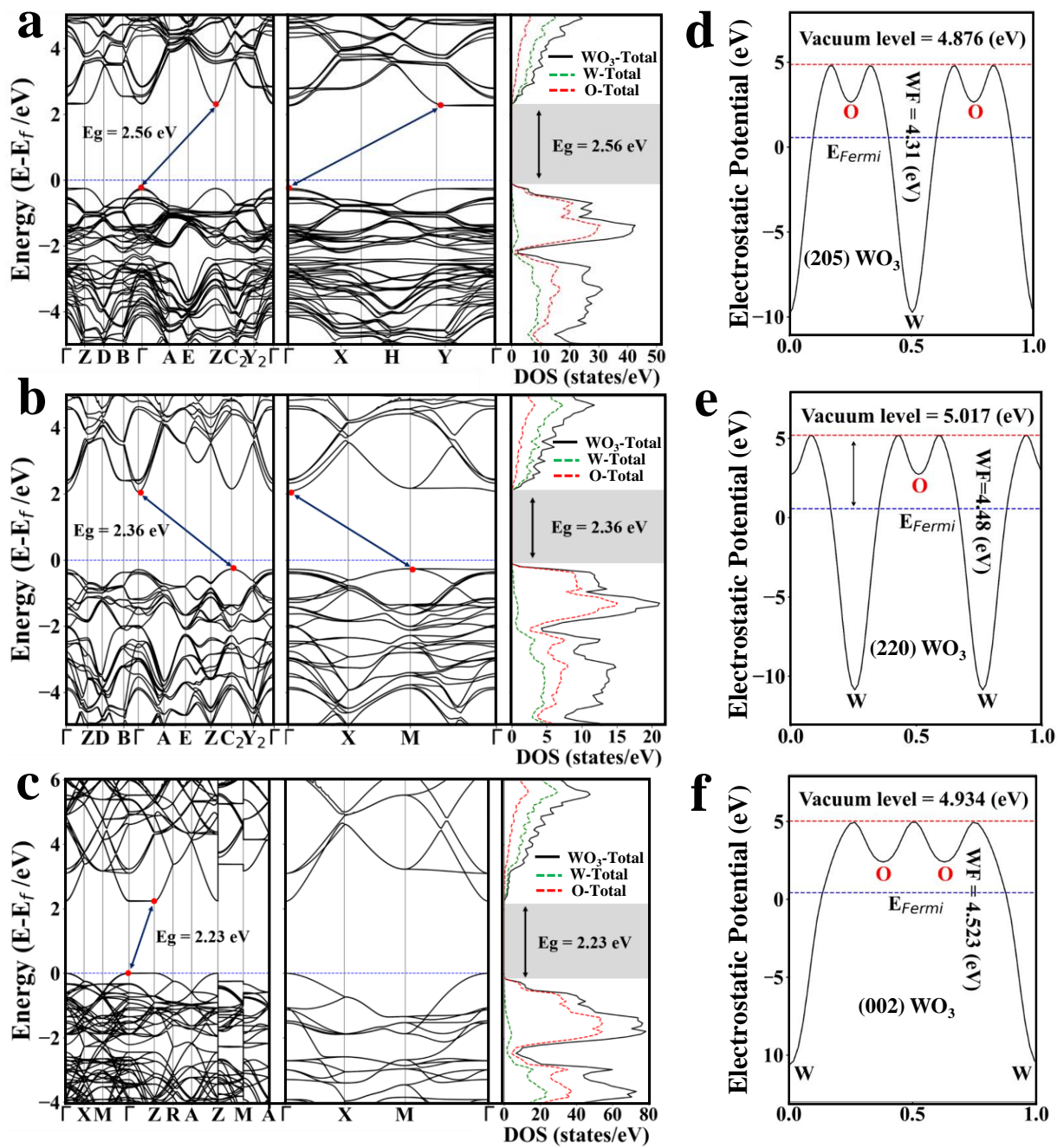

**Figure 4. Density Functional Theory (DFT) calculations for electronic properties.** (a-c) the band structure, TDOS, and (d-f) work function of (205)WO₃, (220)WO₃, and (002)WO₃.

This strong interaction between defect states and intermediates, coupled with favorable electronic properties, underpins the superior photocatalytic performance of (002)WO₃ for CO₂ reduction. We also calculated the work function of each facet to understand how effortlessly electrons can be transferred to the particular facet for effective $CO_2$ reduction. The DFT-calculated work functions of the $(205)WO_3$, $(220)WO_3$, and $(002)WO_3$ facets are 4.31, 4.48, and 4.52 eV versus the electrostatic potential, respectively (**Fig. 4d-f**). As $(002)WO_3$ is dominant and has a more active surface as compared to (220) and $(205)WO_3$. So, here higher work function of $(002)WO_3$ indicates higher electron density or electron cloud as compare to lower work function facets, because the facet having a higher work function can actively participate in $CO_2$ reduction. For further verification of our experimental results, we carried out DFT calculations to observe the $CO_2$ adsorption ability over the prepared catalysts and results revealed the adsorption energy ($E_{ads}$) of $CO_2$ on $(205)WO_3$ to be -2.58, on $(220)WO_3$ to be -2.69, and on $(002)WO_3$ to be -2.72 (**Fig. 5a**). The oxygen vacancies (OVs) in $(002)WO_3$ serve as electron-rich centers that significantly enhance $CO_2$ adsorption and activation. The $CO_2$-TPD data indicate that the stronger adsorption on the (002) facet correlates with a higher OV density, as shown by the desorption peaks in **Fig. 5b**. This suggests that OVs create localized electron-rich sites, promoting the chemisorption of $CO_2$ and subsequent stabilization of intermediates such as COOH and CO [9, 57]. We also carried out Electron Paramagnetic Resonance (EPR) spectroscopy to understand how the unpaired electrons contribute to the redox activity of $WO_3$ facets. In this experiment, we used the 5, 5-dimethyl-l-pyrroline N-oxide (DMPO) to trap the super-oxide radicals ($•O_2^-$) and hydroxyl radicals ($•OH$) evident by the production of the DMPO-$•O_2^-$ (**Fig. 5c**) and DMPO-$•OH$ adducts (**Fig. S23**), respectively. The $(002)WO_3$ was found to be most active for both of these radical adducts owing to the favourable conduction band edge position (-0.42 V vs NHE, PH=7 for DMPO-$•O_2^-$) that is more negative than that of the potential of $O_2/•O_2^-$ potential (-0.33 V vs NHE, PH=7) [58]. Regarding $(205)WO_3$, no signals are found, suggesting no $•O_2^-$ radicals because the CB value (-0.14 V vs NHE, PH=7) is less negative than that of the potential of $O_2/•O_2^-$ potential (-0.33 V vs NHE, PH=7). While the $(220)WO_3$ hardly produces few $•O_2^-$ radicals because the CB value (-0.33 V vs NHE, PH=7) is the same as the potential of $O_2/•O_2^-$ potential (-0.33 V vs NHE, PH=7) and the EPR signals for the DMPO-$•O_2^-$ adducts in $(002)WO_3$ are detected, suggesting several $•O_2^-$ radicals generate because the CB value in this case, the photoelectrons in $(002)WO_3$ have enough reduction potential to form $•O_2^-$ radicals.

On the other hand, as shown in **Figure. S23** no EPR signal can be observed in the $(205)WO_3$, suggesting that the DMPO molecules have no paramagnetic response. For optimized sample (002)WO₃, the intensity peaks of DMPO-•OH radicals as compared to (220)WO₃ is higher. These results indicate that the (002)WO₃ photocatalyst is favourable for producing superoxide radicals (•O₂⁻) and hydroxyl radicals (•OH) because of its favourable CB and VB positions. Additionally, no EPR signal was observed in dark for DMPO-•O₂⁻ and DMPO-•OH radicals as shown in **Fig. S24-25**.

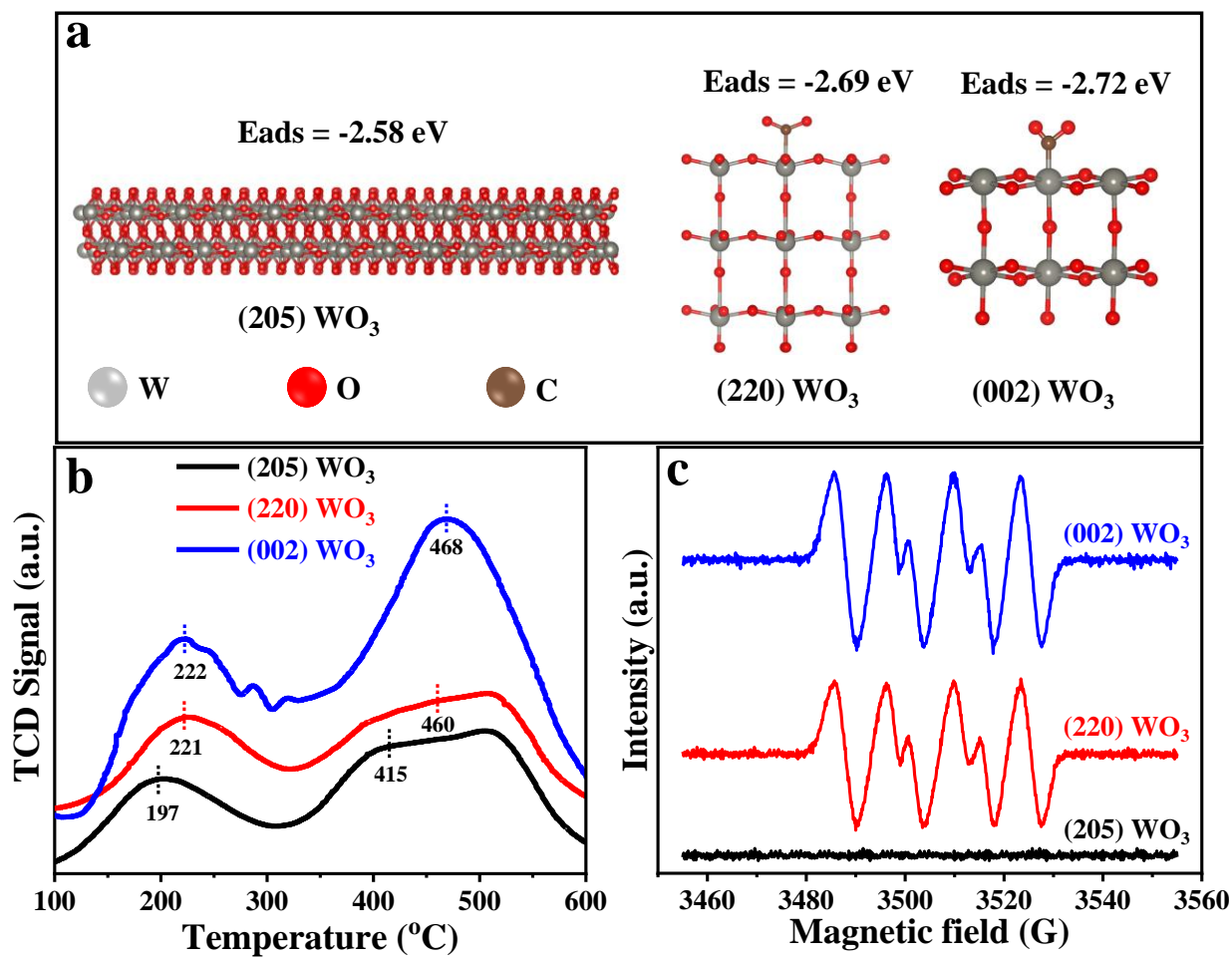

**Figure 5.** Density Functional Theory (DFT) calculations for (a) $CO_2$ adsorption energy, (b) $CO_2$ TPD patterns, and (c) DMPO-trapping EPR spectra for DMPO-•O₂⁻ on (205)WO₃, (220)WO₃, and (002)WO₃.

*In-situ* diffuse reflectance infrared Fourier transform spectroscopy-DRIFTS was used to track the intermediates produced during photocatalytic $CO_2$ reduction on (002)WO₃ in order to study the reaction intermediates. Effective $CO_2$ adsorption and activation on the catalyst surface was demonstrated by the gradual increase of the peaks corresponding to monodentate carbonate (m-$CO_3^{2-}$, 1312 and 1338 cm⁻¹), and bidentate carbonate (b-$CO_3^{2-}$, 1679 cm⁻¹) under light irradiation after 15 minutes, as illustrated in **Fig. 6a**. The increase of the signal at 1698 $cm^{-1}$ during extended illumination indicated that $CO_2^*$ radicals evolved by bending vibrations [59, 60]. The peaks at 1489 and 1508 $cm^{-1}$ correspond to the COOH* intermediate, which is the main intermediate for $CO_2$ reduction to CO (also supported by the DFT reaction pathway, **Fig. 6b-c**) and the small band detected at 2065 $cm^{-1}$ corresponds to CO* that further reveal the production of CO [61, 62], and the peaks intensity was also increased under continuous illumination. This finding supports the CO production in our catalytic system, as illustrated in **Fig. 3i**. Interestingly, CHO*, $CH_3O^*$, and $CH_3^*$ were detected at 1106 $cm^{-1}$, 1208 $cm^{-1}$, and 1361 $cm^{-1}$, respectively, which are important intermediates in the photocatalytic reduction of $CO_2$ to $CH_4$ [59, 63]. The increasing intensities of these peaks under continuous light irradiation validate their roles in the stepwise conversion of $CO_2$ to $CH_4$, which supports the DFT-predicted reaction pathway, and validates the $CH_4$ production seen in the photocatalytic activity as depicted in **Figure 3i**. As a result, as will be discussed below, the discovered intermediates significantly support the DFT predictions in addition to validating the systems photocatalytic activity.

To further validate and gain a deeper understanding of these findings, we employed the Nudged Elastic Band (NEB) method within the framework of Density Functional Theory to identify the transition states along potential reaction pathways and the optimal adsorption configurations, as illustrated in **Fig. 6b**. The adsorption configuration of $WO_3$ is depicted in for each intermediate stage, from the initial $CO_2$ adsorption to the final $CH_4$ production. The C atoms of intermediates like $CO_2^*$, COOH*, CHO*, $CH_2O^*$, and $CH_3^*$ maintain stable bonding with W atoms on the $WO_3$ crystal surface. When $H^+$ and $e^-$ are involved to the process, the removal of the $H_2O$ molecule leads to the breaking of the chemical bond between the oxygen of intermediates and the tungsten atoms. The conversion of $CO_2$ adsorbed on the $WO_3$ surface to COOH* through the hydrogenation process involves a potential energy barrier of 0.5 eV (**Fig. 6c**). This is because electrons localize on the $O_2$ atom of $CO_2$ molecules in the C-O-W adsorption configuration on the surface, aiding in the addition of $H^+$. This process influences the formation energy of COOH* intermediates. Following electron transfer, COOH* decomposes into CO* and $H_2O$. On the other hand, converting COOH* to CO is an endothermic reaction requiring overcoming an energy barrier of 0.2 eV. Direct desorption of CO* species from the surface is also endothermic, whereas further hydrogenation to HCOOH* and CHO*, crucial for $CH_4$ evolution, is thermodynamically favourable. DFT calculations reveal that OVs on the (002) facet reduce the energy barriers for key reaction intermediates. The transition state energies for COOH formation and its subsequent reduction to CO are significantly lower in (002)$WO_3$ compared to other facets (**Fig. 5a** and **Fig.6b-c**). This highlights the crucial role of OVs in enabling selective and efficient $CO_2$ reduction.

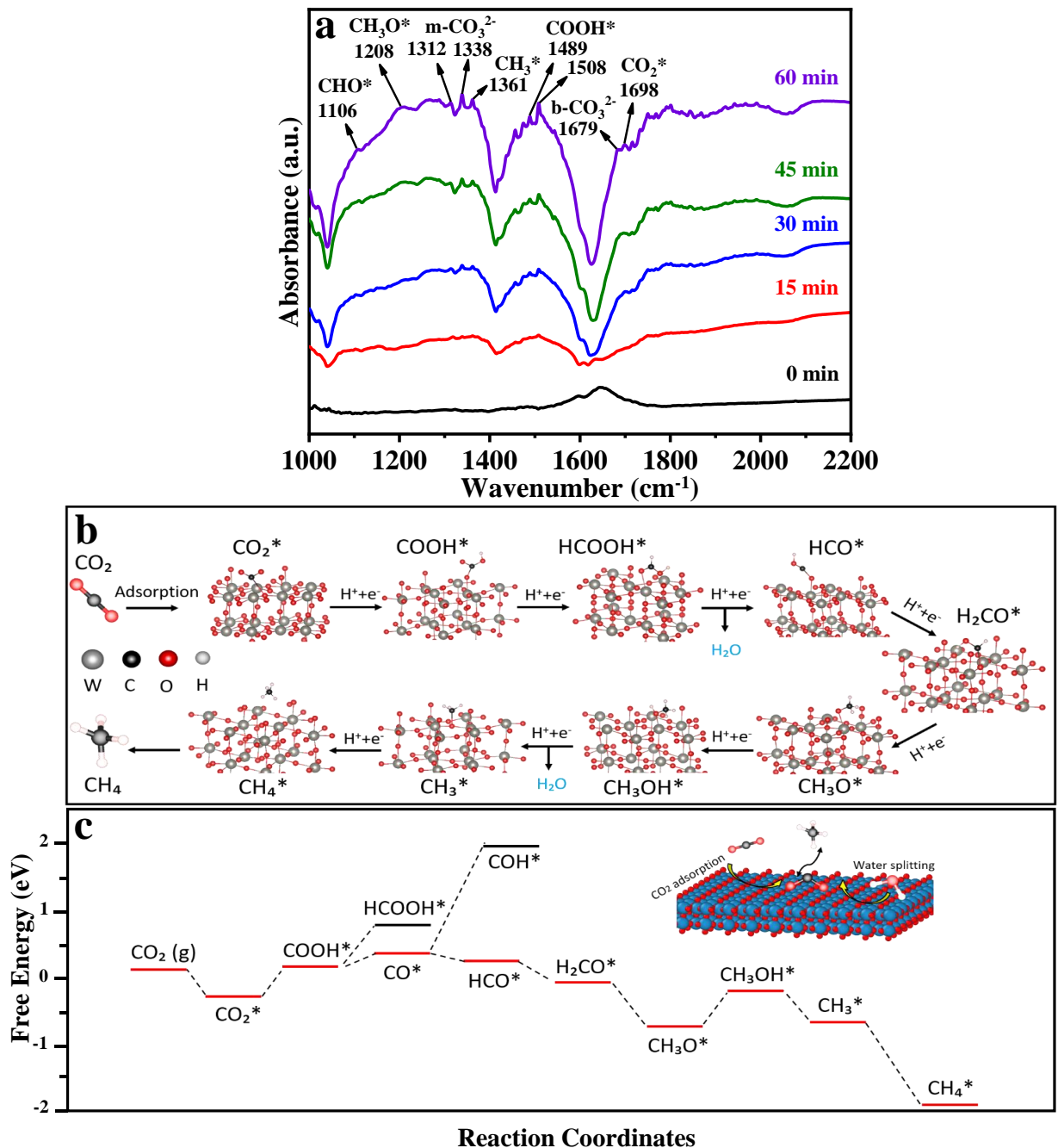

**Figure 6.** (a) in-situ DRIFTS spectra, and (b-c) Reaction pathway of $CO_2$ reduction on (002)$WO_3$.

The generation of $H_2CO*$ intermediates constitute the step that limits the rate of subsequent hydrogenation processes. Nevertheless, the hydrogenation sequence from $H_2CO*$ to $H_3CO*$ and ultimately to $CH_4$ demonstrates thermodynamic preference, thereby enabling $WO_3$ to achieve high selectivity for $CH_4$ production. It can be seen that the pathway from $CO*$ to $CH_4$ through intermediates such as $HCO*$, $H_2CO*$, $CH_3O*$, and $CH_3*$ exhibits a sequential decrease in free energy. Consequently, it is important to highlight that desorption of $H_2CO*$ and the formation of $CH_3OH*$ from $CH_3O*$ were both observed to be highly endothermic processes, indicating that the preferential pathway for $CH_4$ evolution aligns well with the minimal detection of CO. Moreover, the free energy ($\Delta G$) at 0 V vs. NHE for adsorbate states, computed at room temperature on the 002 surface of the $WO_3$ structure and NEB convergence across the reaction pathway are also shown in **Table. S1** and **Fig. S26**.

## Conclusions

In conclusion, this study demonstrates a novel approach to enhancing the photocatalytic properties of $WO_3$ by inducing topological modifications through heat treatment. The creation of oxygen-vacant more-porous nanosheets with exposed active (002) facets leads to superior electron conduction and increased reduction potential, making them highly effective for $CO_2$ reduction. The findings highlight the significant performance improvements of these modified $WO_3$ nanosheets compared to their less-porous counterparts and bulk $WO_3$. This research not only advances our understanding of atomic structure modulation in $WO_3$ but may also open new avenues for applications in optoelectronics, sensors, and energy conversion.

## Declaration of competing interest

The authors declare that they have no known competing financial interests or personal relationships that could have appeared to influence the work reported in this paper.

## Acknowledgments

This work was supported by the National Natural Science Foundation of China (Grant Nos.

22350410385, 22150610469, 11774044, and 52072059). Author ANA acknowledges Researchers Supporting Project Number (RSP2025R304), King Saud University, Riyadh, Saudi Arabia.

## References

[1]     Lenton, T. M.; Rockström, J.; Gaffney, O.; Rahmstorf, S.; Richardson, K.; Steffen, W.; Schellnhuber, H. J., Climate tipping points — too risky to bet against. Nature. **2019**, 575, 52.

[2]     Ismail, P. M.; Ali, S.; Ali, S.; Li, J.; Liu, M.; Yan, D.; Raziq, F.; Wahid, F.; Li, G.; Yuan, S., Photoelectron "bridge" in Van Der Waals heterojunction for enhanced    photocatalytic $CO_2$ conversion under visible light. Adv. Mater. **2023**, 35, 2303047.

[3]     Davis, S. J.; Lewis, N. S.; Shaner, M.; Aggarwal, S.; Arent, D.; Azevedo, I. L.; Benson, S. M.; Bradley, T.; Brouwer, J.; Chiang, Y. M.; Clack, C. T. M.; Cohen, A.; Doig, S.; Edmonds, J.; Fennell, P.; Field, C. B.; Hannegan, B.; Hodge, B. M.; Hoffert, M. I.; Ingersoll, E.; Jaramillo, P.; Lackner, K. S.; Mach, K. J.; Mastrandrea, M.; Ogden, J.; Peterson, P. F.; Sanchez, D. L.; Sperling, D.; Stagner, J.; Trancik, J. E.; Yang, C. J.; Caldeira, K., Net-zero emissions energy systems. Science. **2018**, 360, 1419.

[4]     Raziq, F.; Rahman, M. Z.; Ali, S.; Ali, R.; Ali, S.; Zada, A.; Wu, X.; Gascon, J.; Wang, Q.; Qiao, L., Enhancing Z-scheme photocatalytic $CO_2$ methanation at extended visible light (>600 nm): Insight into charge transport and surface catalytic reaction mechanisms. Chem. Eng. J. **2024**, 479, 147712.

[5]     Ismail, P. M.; Ali, S.; Raziq, F.; Bououdina, M.; Abu-Farsakh, H.; Xia, P.; Wu, X.; Xiao, H.; Ali, S.; Qiao, L., Stable and robust single transition metal atom catalyst for $CO_2$ reduction supported on defective $WS_2$. Appl. Surf. Sci. **2023**, 624, 157073.

[6]     Loh, J. Y. Y.; Kherani, N. P.; Ozin, G. A., Persistent $CO_2$ photocatalysis for solar fuels in the dark. Nat. Sustain. **2021**, 4, 466-473.

[7]     Rahman, M. Z.; Gascon, J., Determining the sequence of charge transport events and their roles on the limit of quantum efficiency of photocatalytic hydrogen production. Matter. **2023**, 6, 2081–2093.

[8]     Rahman, M. Z.; Raziq, F.; Zhang, H.; Gascon, J., Key Strategies for Enhancing $H_2$ Production in Transition Metal Oxide Based Photocatalysts. Angew. Chem. Int. Ed. **2023**, 62, 48.

[9]     Ali, S.; Ali, S.; Khan, I.; Zahid, M.; Muhammad Ismail, P.; Ismail, A.; Zada, A.; Ullah, R.; Hayat, S.; Ali, H.; Kamal, M. R.; Alibrahim, K. A.; Bououdina, M.; Hasnain Bakhtiar, S.; Wu, X.; Wang, Q.; Raziq, F.; Qiao, L., Molecular modulation of interfaces in a Z-scheme van der Waals heterojunction for highly efficient photocatalytic $CO_2$ reduction. J. Colloid Interface Sci. **2024**, 663, 31-42.

[10]    Collado, L.; Reñones, P.; Fermoso, J.; Fresno, F.; Garrido, L.; Pérez-Dieste, V.; Escudero, C.; Hernández-Alonso, M. D.; Coronado, J. M.; Serrano, D. P., The role of the surface acidic/basic centers and redox sites on $TiO_2$ in the photocatalytic $CO_2$ reduction. Appl. Catal. B: Environ. **2022**, 303, 120931.

[11]    Wu, X.; Wang, C.; Wei, Y.; Xiong, J.; Zhao, Y.; Zhao, Z.; Liu, J.; Li, J., Multifunctional photocatalysts of Pt-decorated 3DOM perovskite-type $SrTiO_3$ with enhanced $CO_2$

adsorption and photoelectron enrichment for selective $CO_2$ reduction with $H_2O$ to $CH_4$. J. Catal. **2019**, 377, 309-321.

[12]  Khalid, N.; Ishtiaq, H.; Ali, F.; Tahir, M.; Naeem, S.; Ul-Hamid, A.; Ikram, M.; Iqbal, T.; Kamal, M. R.; Alrobei, H., Synergistic effects of Bi and N doped on ZnO nanorods for efficient photocatalysis. Mater. Chem. Phys. **2022**, 289, 126423.

[13]  Rahman, M. Z.; Edvinsson, T.; Gascon, J., Hole utilization in solar hydrogen production. Nat. Rev. Chem. **2022**, 6, 243-245.

[14]  Rahman, M. Z.; Zhang, J.; Tang, Y.; Davey, K.; Qiao, S.-Z., Graphene oxide coupled carbon nitride homo-heterojunction photocatalyst for enhanced hydrogen production. Mater. Chem. Front. **2017**, 1, 562-571.

[15]  Ali, H.; Liu, M.; Ali, S.; Ali, A.; Ismail, P. M.; Ullah, R.; Ali, S.; Raziq, F.; Bououdina, M.; Hayat, S.; Ali, U.; Zhou, Y.; Wu, X.; Zhong, L.; Zhu, L.; Xiao, H.; Xia, P.; Qiao, L., Constructing copper Phthalocyanine/Molybdenum disulfide (CuPc/MoS$_2$) S-scheme heterojunction with S-rich vacancies for enhanced Visible-Light photocatalytic $CO_2$ reduction. J. Colloid Interface Sci. **2024**, 665, 500-509.

[16]  Liu, Y.; Liang, L.; Xiao, C.; Hua, X.; Li, Z.; Pan, B.; Xie, Y. J., Promoting photogenerated holes utilization in pore-rich $WO_3$ ultrathin nanosheets for efficient oxygen-evolving photoanode. Adv. Energy Mater. **2016**, 6, 1600437.

[17]  Mendieta-Reyes, N. E.; Díaz-García, A. K.; Gómez, R., Simultaneous electrocatalytic $CO_2$ reduction and enhanced electrochromic effect at $WO_3$ nanostructured electrodes in acetonitrile. ACS Catal. **2018**, 8, 1903-1912.

[18]  Liao, M.; Su, L.; Deng, Y.; Xiong, S.; Tang, R.; Wu, Z.; Ding, C.; Yang, L.; Gong, D., Strategies to improve $WO_3$-based photocatalysts for wastewater treatment: a review. J. Mater. Sci. **2021**, 56, 14416-14447.

[19]  Huang, S.; Long, Y.; Ruan, S.; Zeng, Y.-J., Enhanced photocatalytic $CO_2$ reduction in defect-engineered Z-scheme $WO_3–x/g-C_3N_4$ heterostructures. ACS Omega. **2019**, 4, 15593-15599.

[20]  Sun, S.; Watanabe, M.; Wu, J.; An, Q.; Ishihara, T. J., Ultrathin $WO_3 \cdot 0.33 H_2O$ nanotubes for $CO_2$ photoreduction to acetate with high selectivity. J. Am. Chem. Soc. **2018**, 140, 6474-6482.

[21]  Li, B.; Sun, L.; Bian, J.; Sun, N.; Sun, J.; Chen, L.; Li, Z.; Jing, L., Controlled synthesis of novel Z-scheme iron phthalocyanine/porous $WO_3$ nanocomposites as efficient photocatalysts for $CO_2$ reduction. Appl. Catal. B: Environ. **2020**, 270, 118849.

[22]  Li, X.; Song, X.; Ma, C.; Cheng, Y.; Shen, D.; Zhang, S.; Liu, W.; Huo, P.; Wang, H., Direct Z-scheme $WO_3$/graphitic carbon nitride nanocomposites for the photoreduction of $CO_2$. Nano Mater. **2020**, 3, 1298-1306.

[23]  Lei, B.; Cui, W.; Chen, P.; Chen, L.; Li, J.; Dong, F., C–doping induced oxygen-vacancy in $WO_3$ nanosheets for $CO_2$ activation and photoreduction. ACS Catal. **2022**, 12, 9670-9678.

[24]  Yang, G.; Xiong, J.; Lu, M.; Wang, W.; Wen, Z.; Li, S.; Li, W.; Wen, Z.; Li, S.; Chen R.; Cheng, G., Co-embedding oxygen vacancy and copper particles into titanium-based oxides ($TiO_2$, $BaTiO_3$, and $SrTiO_3$) nanoassembly for enhanced $CO_2$ photoreduction through surface/interface synergy. J. Colloid Interface Sci. **2022**, 624, 348-361.

[25]  Tong, X.; W. Chen.; G. Cheng., Oxygen vacancy-enrich Ag/brookite $TiO_2$ Schottky junction for enhanced photocatalytic $CO_2$ reduction. J. Mol. Catal. **2024**, 560, 114140.

[26]     Xiong, J.; Zhang, M.; Lu M.; K Zhao.; Han C.; Cheng G.; Wen Z., Achieving simultaneous Cu particles anchoring in meso-porous $TiO_2$ nano-fabrication for enhancing photo-catalytic $CO_2$ reduction through rapid charge separation. Chin. Chem. Lett. **2022**, 33, 1313-1316.

[27]     Yang, G.; Qiu, P.; Xiong J.; Zhu X.; Cheng G., Facilely anchoring $Cu_2O$ nanoparticles on mesoporous $TiO_2$ nanorods forenhanced photocatalytic $CO_2$ reduction through efficient charge transfer. Chin. Chem. Lett. **2022**, 33, 3709-3712.

[28]     Pan, L.; Dai, L.; Burton, O. J.; Chen, L.; Andrei, V.; Zhang, Y.; Ren, D.; Cheng, J.; Wu, L.; Frohna, K.; Abfalterer, A.; Yang, T. C.; Niu, W.; Xia, M.; Hofmann, S.; Dyson, P. J.; Reisner, E.; Sirringhaus, H.; Luo, J.; Hagfeldt, A.; Gratzel, M.; Stranks, S. D., High carrier mobility along the [111] orientation in $Cu_2O$ photoelectrodes. Nature. **2024**, 628, 765-770.

[29]     Wagner, A.; Sahm, C. D.; Reisner, E., Towards molecular understanding of local chemical environment effects in electro- and photocatalytic $CO_2$ reduction. Nat. Catal. **2020**, 3, 775-786.

[30]     Moss, B.; Wang, Q.; Butler, K. T.; Grau-Crespo, R.; Selim, S.; Regoutz, A.; Hisatomi, T.; Godin, R.; Payne, D. J.; Kafizas, A.; Domen, K.; Steier, L.; Durrant, J. R., Linking in situ charge accumulation to electronic structure in doped $SrTiO_3$ reveals design principles for hydrogen-evolving photocatalysts. Nat. Mater. **2021**, 20, 511-517.

[31]     Takata, T.; Jiang, J.; Sakata, Y.; Nakabayashi, M.; Shibata, N.; Nandal, V.; Seki, K.; Hisatomi, T.; Domen, K., Photocatalytic water splitting with a quantum efficiency of almost unity. Nature. **2020**, 581, 411-414.

[32]     Liu, J.; Xu, S.-M.; Li, Y.; Zhang, R.; Shao, M., Facet engineering of $WO_3$ arrays toward highly efficient and stable photoelectrochemical hydrogen generation from natural seawater. Appl. Catal. B: Environ. **2020**, 264, 118540.

[33]     Zhang, Y.; Xia, B.; Ran, J.;Davey, K.; Qiao, S. Z., Atomic-level reactive sites for semiconductor-based photocatalytic $CO_2$ reduction. Adv. Energy Mater. **2020**, 10, 1903879.

[34]     Sun, S.; He, L.; Yang, M.; Cui, J.; Liang, S., Facet junction engineering for photocatalysis: a comprehensive review on elementary knowledge, facet-synergistic mechanisms, functional modifications, and future perspectives. Adv. Funct. Mater. **2022**, 32, 2106982.

[35]     Xiong, Z.; Lei, Z.; Li, Y.; Dong, L.; Zhao, Y.; Zhang, J., A review on modification of facet-engineered $TiO_2$ for photocatalytic $CO_2$ reduction. J. Photochem. **2018**, 36, 24-47.

[36]     Tan, C.; Zhang, H., Wet-chemical synthesis and applications of non-layer structured two-dimensional nanomaterials. Nat. Commun. **2015**, 6, 7873.

[37]     Chhowalla, M.; Shin, H. S.; Eda, G.; Li, L. J.; Loh, K. P.; Zhang, H., The chemistry of two-dimensional layered transition metal dichalcogenide nanosheets. Nat. Chem. **2013**, 5, 263-275.

[38]     Paul, A.; Ghosh, S.; Kolya, H.; Kang, C.-W.; Murmu, N. C.; Kuila, T., New insight into the effect of oxygen vacancies on electrochemical performance of nickel-tin oxide/reduced graphene oxide composite for asymmetric supercapacitor. J. Energy Storage. **2023**, 62, 106922.

[39]     Li, J.-J.; Zhang, M.; Weng, B.; Chen, X.; Chen, J.; Jia, H.-P., Oxygen vacancies mediated charge separation and collection in $Pt/WO_3$ nanosheets for enhanced photocatalytic performance. Appl. Surf. Sci. **2020**, 507, 145133.

[40]     Rao, F.; An, Y.; Zhu, G.; Gong, S.;Zhu, L.; Lu, H.; Shi, X.; Huang, Y.; Zhang, F.; Hojamberdiev, M., Unveiling the effects of facet-dependent oxygen vacancy on $CeO_2$ for

electron structure and surface intermediates in $CO_2$ photoreduction reaction. Sep. Purif. Technol. **2024**, 333, 125951.

[41]  Li, Q.; Guo, B.; Yu, J.; Ran, J.; Zhang, B.; Yan, H.; Gong, J. R., Highly Efficient Visible-Light-Driven Photocatalytic Hydrogen Production of CdS-Cluster-Decorated Graphene Nanosheets. J. Am. Chem. Soc. **2011**, 133, 10878-10884.

[42]  Zhu, X.; Zhang, Y.; Wang, Y.; Liu, Y.; Wu, Z., Oxygen-deficient $WO_3$ for stable visible-light photocatalytic degradation ofacetaldehyde within a wide humidity range. Chem. Eng. J. **2024**, 491, 152193.

[43]  Yan, L.; Dong, G.; Huang, X.; Zhang, Y.; Bi, Y., Unraveling oxygen vacancy changes of $WO_3$ photoanodes for promoting oxygen evolution reaction. Appl. Catal. B: Environ. **2024**, 345, 123682.

[44]  Shi, K.; Wang, F.; Li, X.; Huang, W.; Lu, K.; Yu, C.; Yang, K., Defect-engineered $WO_{3-x}$ nanosheets for optimized photocatalytic nitrogen fixation and hydrogen production. J. Mater. Sci. **2023**, 58,16309-16321.

[45]  Li, H.; Shen, Q.; Zhang, H.; Gao, J.; Jia, H.; Liu, X.; Li, Q.; Xue, J., Oxygen vacancy-mediated $WO_3$ phase junction to steering photogenerated charge separation for enhanced water splitting. J. Adv. Ceram. **2022**, 11, 1873-1888.

[46]  Liu, R.; Shi, Y.; Lin, L.; Wang, Z.; Liu, C.; Bi, J.; Hou, Y.; Lin, S.; Wu, L., Surface Lewis acid sites and oxygen vacancies of $Bi_2WO_6$ synergistically promoted photocatalytic degradation of levofloxacin. Appl. Surf. Sci. **2022**, 605, 154822.

[47]  Liu, L.; Liu, J.; Sun, K.; Wan, J.; Fu, F.; Fan, J., Novel phosphorus-doped $Bi_2WO_6$ monolayer with oxygen vacancies for superior photocatalytic water detoxication and nitrogen fixation performance. Chem. Eng. J. **2021**, 411, 128629.

[48]  Q, Liu.; Wang, F.; Lin, H.; Xie, Y.; Tong, N.; Lin, J.; Zhang, Z.; Wang, X., Surface oxygen vacancy and defect engineering of $WO_3$ for improved visible light photocatalytic performance. Catal. Sci. Technol. **2018**, 17, 4399-4406.

[49]  Yi, H.; Yan, M.; Huang, D.; Wang, H.; Shen, M., Synergistic effect of artificial enzyme and 2D nano-structured $Bi_2WO_6$ for eco-friendly and efficient biomimetic photocatalysis. Appl. Catal. B: Environ. **2019**, 250, 52-62.

[50]  Gao, W.; Li, G.; Wang, Q.; Zhang, L.; Xia Y., Ultrathin porous $Bi_2WO_6$ with rich oxygen vacancies for promoted adsorption-photocatalytic tetracycline degradation. Chem. Eng. J. **2023**, 464, 142694.

[51]  Rahman, M. Z.; Maity, P.; Mohammed, O. F.; Gascon, J., Insight into the role of reduced graphene oxide for enhancing photocatalytic hydrogen evolution in disordered carbon nitride. Phys. Chem. Chem. Phys. **2022**, 24, 11213-11221.

[52]  Rahman, M.; Davey, K., Enabling Pt-free photocatalytic hydrogen evolution on polymeric melon: Role of amorphization for overcoming the limiting factors. Phys. Rev. Mater. **2018**, 2, 125402.

[53]  Rahman, M. Z.; Batmunkh, M.; Bat-Erdene, M.; Shapter, J. G.; Mullins, C. B., p-Type BP nanosheet photocatalyst with AQE of 3.9% in the absence of a noble metal cocatalyst: investigation and elucidation of photophysical properties. J. Mater. Chem. A. **2018**, 6, 18403-18408.

[54]  Rahman, M. Z.; Tang, Y.; Kwong, P., Reduced recombination and low-resistive transport of electrons for photo-redox reactions in metal-free hybrid photocatalyst. Appl. Phys. Lett. **2018**, 112, 253902.

[55]  Rahman, M. Z.; Moffatt, J.; Spooner, N., Topological carbon nitride: localized photon

absorption and delocalized charge carrier separation at intertwined photocatalyst interfaces. Mater. Horiz. **2018**, 5, 553-559.

[56]   Li, H.; Shang, J.; Yang, Z.; Shen, W.; Ai, Z.; Zhang, L., Oxygen Vacancy Associated Surface Fenton Chemistry: Surface Structure Dependent Hydroxyl Radicals Generation and Substrate Dependent Reactivity. Environ. Sci. Tech. **2017**, 51, 5685-5694.

[57]   Zeng, D.; Wang, H.; Zhu, X.; Cao, H.; Zhou, Y.; Wang, W.; Zhang, L.; Wang, W., Single-atom copper modified hexagonal tungsten oxide for efficient photocatalytic $CO_2$ reduction to acetic acid. Chem. Eng. J. **2023**, 451, 138801.

[58]   Li, Z.; Hou, J.; Zhang, B.; Cao, S.; Wu, Y.; Gao, Z.; Nie, X.; Sun, L., Two-dimensional Janus heterostructures for superior Z-scheme photocatalytic water splitting. Nano Energy. **2019**, 59, 537-544.

[59]   Z, Li.; Rong, Y.; Liang, J.;  Li, Z.; Liang, Z.; Hou, Y., In-situ generation of $Bi^0$ NCs and vacancies on Bi-CTS/BiOBr heterostructures accelerate electron transfer for promoting photocatalytic reduction of  $CO_2$. J. Environ. Chem. J. **2022**, 10, 108819.

[60]   B, Lei.; W. Cui, P.; Chen, L. Chen.;  J. Li.;  F. Dong., C–Doping Induced Oxygen-Vacancy in $WO_3$ Nanosheets for $CO_2$ Activation and Photoreduction. ACS Catal. **2022**,  12, 9670-9678.

[61]   Y, He.; Zhang, Z.; Yu, J.; Xu, D.; Liu, C.; Pan, Y.; Macyk, W.; Xu., Selective conversion of $CO_2$ to $CH_4$ enhanced by $WO_3$/$In_2O_3$ S-scheme heterojunction photocatalysts with efficient $CO_2$ activation. J. Mater. Chem. A. **2023**, 11, 14860-14869.

[62]   G, Chen.; Zhou, Z.; Li, B.; Lin, X.; Yang, C.; Lin, X.; Yang, C.; Fang, Y.; Lin, W.; Hou, Y.; Zhang, G.; Wang, S., S-scheme heterojunction of crystalline carbon nitride nano-sheets and ultrafine $WO_3$ nanoparticles for photocatalytic $CO_2$ reduction. J. Environ. Sci. **2024**, 140, 103-112.

[63]   M, Zahid.; Ismail, A.; Ullah, R.; Ali, U.; Raziq, F.; Alrebi, TA.; Alodhyayb, AN.; Ali, S.; Qiao, L.,  Pt-N catalytic centres concisely enhance interfacial charge transfer in amines functionalized Pt@MOFs for selective conversion of $CO_2$ to $CH_4$. J. Colloid Interface Sci. **2024**, 672, 370-382.